# Energy Efficient Massive MIMO Array Configurations

Hardy Halbauer, Andreas Weber, Dirk Wiegner, Thorsten Wild
Nokia Bell Labs
Stuttgart, Germany
{firstname.lastname}@nokia-bell-labs.com

*Abstract*—The high spectral efficiency of massive MIMO (Multiple Input Multiple Output) is mainly achieved through the exploitation of spatial multiplexing, i.e. by using a high number of MIMO layers that are applied simultaneously to many users. The power consumption of a massive MIMO base station is determined by the hardware driving a high number of antenna ports and elements. This paper focuses on practical deployment situations with varying user load. During hours with low number of users a certain significant part of hardware power consumption would remain with conventional massive MIMO processing, while the full potential of spectral efficiency cannot be exploited due to the low number of users, resulting in low power efficiency and cost. We investigate the impact of different hybrid array architectures on spectral efficiency, average user throughput and power consumption and show how to design a massive MIMO system with significantly improved energy efficiency for a given target scenario, while maintaining a targeted service quality.

I. INTRODUCTION

Upcoming 5G systems promise a significantly improved system throughput and service quality. This may be achieved by simply increasing the carrier bandwidth, by aggregation of carriers, or by increasing the density of base stations, which are rather expensive solutions. A further solution is the usage of a higher number of Single User MIMO (SU-MIMO) layers. However, this means to put high complexity into the User Equipment (UE). More attractive is the re-use of resources by Multi User MIMO (MU-MIMO). Following this solution, we put the complexity into the base station, by usage of antenna arrays with a large number of antenna elements, called Massive MIMO. The UEs can be very simple; however, we can still combine MU-MIMO and SU-MIMO when UEs are equipped with multiple antenna elements.

Good channel knowledge is key to be able to separate MIMO layers and consequently, to achieve high transmission quality. Massive MIMO systems are beneficially deployed as a TDD (Time Division Duplex) system with uplink (UL) pilot based channel sounding. Due to uplink/downlink channel reciprocity UL channel measurements can be used for the downlink (DL). In this case, calibrated antenna systems are required. Pilot contamination [1] by neighbor cell UEs leads to distorted UL channel measurements and therefore to lower SINR in the downlink. This can be reduced by higher UL pilot re-use factors and by using UL pilot power control.

Energy-Efficiency (EE), measured e.g. in Joule/bit is an extremely important requirement by mobile system operators, as it has direct impact on OPEX (operating expense) cost. However, it is evenly important to maintain a good user perception in terms of service quality, peak and average data rates.

The basic idea of massive MIMO was first published in [1] which describes all fundamental challenges. [2] presents a detailed overview of massive MIMO and introduces the aspect of EE.

Energy-Efficiency for massive MIMO has been tackled by various publications [3] - [9]. These papers already use an enhanced power consumption model that considers not only the impact of power amplifiers, but also the contributions of the rest of the transmit path, e.g. from baseband processing, DAC (Digital Analog Converter), filters, etc. They show that the optimum number of transmit (TX) antennas is reduced when optimizing with respect to EE using a detailed energy model. In [4] the authors show that this is also true, when introducing hardware impairments.

In [10], in addition to mathematical analysis, a simulative study has been performed, based on a 3D channel model [12]. However, the impact of the channel characteristic on selecting antenna elements has not been discussed.

In contrast to these studies, which target at high energy efficiency for a specific system configuration and user density, we follow a more practical issue: How to configure and operate a massive MIMO system under strong variation of number of users in an energy efficient way. The high spectral efficiency of massive MIMO systems is achieved by a high number of simultaneously transmitted MIMO layers. Therefore, making full use of spectral efficiency requires spatial multiplexing, realized with a high number of antenna ports and antenna elements. This is beneficial only if a high number of UEs can be served simultaneously. For a lower number of UEs (which in practical deployments regularly occurs e.g. during night time) the potential of a large massive MIMO antenna array cannot be fully exploited, due to lack of multiplexing possibilities. Various massive MIMO RF (Radio Frequency) and antenna architectures have been discussed in literature (full digital, hybrid fully connected or hybrid subarray structures, [11]). To have maximum degrees of freedom for MIMO precoding, individual transmit and receive chains per antenna element are the best choice for performance. But the hardware and processing effort scales with the number of antenna elements and not with the number of users. Therefore, the power consumption (and in consequence the OPEX cost for the operator) of such a system remains high, even with a low number of active users in the system.

Therefore, we propose to maintain the Quality of Service (QoS) under different traffic load situations by adaptation of the antenna array. This is achieved by switching off parts of the antenna array during low traffic load, while maintaining the average user throughput and, hence, using the antenna array with a higher energy efficiency. We assess the impact of different array architectures and dimensioning on DL performance. We study the performance of realizable hardware structures and antenna arrays based on realistic channel models. In parallel, the impact on power consumption is evaluated, based on a realistic power consumption model with focus on analog frontend components and power amplifier efficiency. In relation to the analog front-end, we do not expect a high impact of the baseband processing power consumption, that we neglect for this study. The power consumption versus performance trade-off for different deployment situations and massive MIMO architectures is highlighted. Finally, an energy efficient massive MIMO system design, using adaptation of the antenna and RF architecture, is proposed and discussed.

The paper is organized as follows: Section II presents an overview of the investigated antenna array architectures. Section III presents the investigated system model, simulation assumptions and performance results. Section IV depicts the power consumption model and merges performance and power consumption results into our findings on power efficiency. Finally, in Section V we close the paper with conclusions and outlook.

## II. ANTENNA ARRAY ARCHITECTURES

The straightforward approach to realize a massive MIMO system is to control each antenna element individually by a precoder, which maps a number of $K$ MIMO layers $s = [\ s_1\ ...\ s_K]^T$ to $N$ antenna port signals $t = [t_1\ ...\ t_N]^T$ forming an array with $N$ antenna elements (Fig. 1). The signal format considered here is an OFDM (Orthogonal Frequency Division Multiplexing) signal with 15 kHz subcarrier spacing and bandwidths between 10 MHz and 100 MHz, as specified by 3GPP for LTE and New Radio systems [13]. This allows the most flexible precoding to fully exploit MIMO capability. However, this flexibility requires a separate RF conversion chain and power amplifier (PA) per antenna element, leading to a high hardware effort.

Hybrid array structures are mapping $K$ MIMO layers $s$ to a number $P$ of antenna port signals $t = [t_1\ ...\ t_P]^T$, which is much lower than the number of antenna elements $N$. An overview on hybrid array types can be found e.g. in [11]. The number of analog-to-digital and digital-to-analog converters, RF modules and PAs is reduced by a factor of $N/P$, which is advantageous especially in the millimeter-wave frequency bands because of the higher cost of mm-wave components. In contrast, our investigations focus on power consumption and energy efficiency in the frequency band below 6 GHz. We consider the hybrid "subarray" structure, with each antenna port connected to a separate subarray, as a low complexity variant (Fig. 2). For our investigations the analog precoder is fixed and represents a feeder network for the subarray antenna elements.

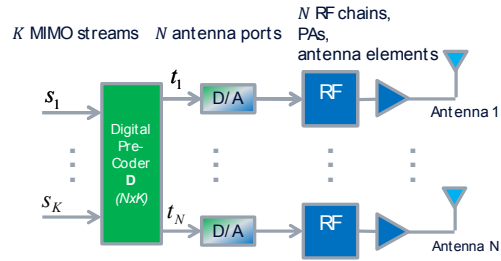

Fig. 1. Full digital antenna array

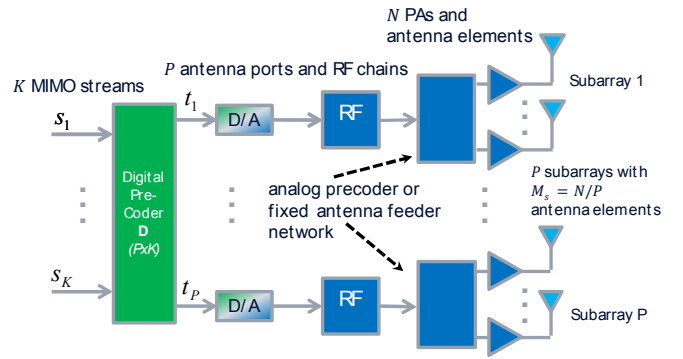

Fig. 2. Hybrid subarray antenna architecture

The combination of the main parameters $K$, $P$ and $N$, together with the array size given in terms of wavelengths, are influencing the spectral efficiency and power consumption of a specific deployment scenario. $P$ is an indicator for the precoding flexibility and limits the number of simultaneous UEs to be served. $N$ in combination with the array size influences the spatial diversity or separation of individual MIMO layers.

In our investigations we compare the achievable performance of the different array types shown in Fig. 3. Array type A comprises $N = 256$ elements, grouped into 32 subarrays with eight cross-polarized elements each. $M_s = 4$ elements of the same polarization represent one antenna port, so that in total $P = 64$ antenna ports are available and the cross-polarized ports are co-located. The relation between $P$, $N$ and the subarray size $M_s$ is $P = N/M_s$.

Array types A, E and F use identical number of ports $P = 64$. The only difference is the number of elements per subarray $M_s$ (4, 2 and 1, respectively) and therefore the total number of elements $N = 256, 128$ and $64$. Array types K and L have the same $N$ as types E and F, respectively, but the subarray size is $M_s = 4$ in all cases. This leads to a reduction of ports to $P = 32$ and $16$, respectively.

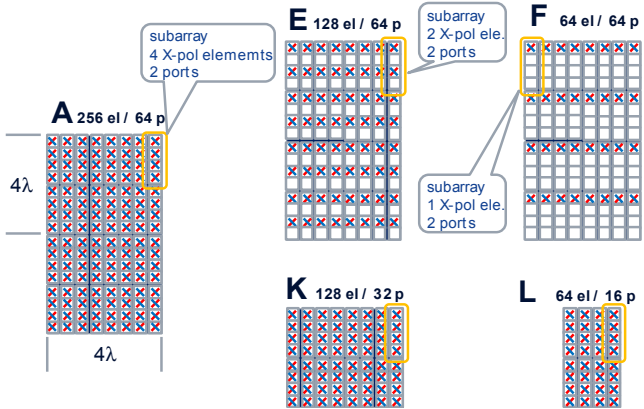

Fig. 3. Different investigated array types A, E, F, K, L with 256, 128 and 64 elements, supporting 64, 32 and 16 ports (el = antenna elements, p = ports)

## III. SYSTEM SIMULATION

### A. System model and parameters

The most important simulation assumptions are listed in TABLE I. Our simulations are performed with a 3GPP calibrated system level simulator. We use a 3D channel model for Urban Macro (UMa) published by 3GPP [12]. Furthermore, we use the settings proposed in [12] except for the number of cells which is set to one central cell surrounded by one tier of further cells (7x3 sectors in total). We apply wrap around, i.e. each cell behaves like a center cell with respect to extra cell interference. We assume a TDD system with channel reciprocity, whereby the UL is purely used for channel measurements (see Fig. 4). UEs transmit one reference pilot signal per coherence block. For orthogonality between all pilots available in a cell, we assume that the pilot sequence length corresponds to the number of users located in a cell (3 sectors) multiplied with the number of antennas per UE (2 in the considered case). The pilot sequences are repeated in every cell. Consequently, there is no pilot contamination between the sectors of a cell, but between different cells, which is accounted for.

The UEs perform UL power control for the transmission of pilots. We assume that 90% of the users can fully compensate the coupling loss (which considers propagation loss plus shadowing minus BS antenna element gain). The 10% UEs with the highest coupling loss transmit their pilots with full power. The UL power control coverage of 90% turned out to be optimal for the considered UMa scenario.

For scheduling we use SUS (Semi-Orthogonal User Selection) [14] and single layer transmission. SUS maximizes the expected throughput based on a greedy algorithm. After scheduling decision, Zero Forcing (ZF) is used for the precoder, which is calculated based on the interfered UL pilot measurements of the scheduled set of UEs. During user selection and for precoding we assume equal power allocation per scheduled UE. For this study we assume perfect link adaptation based on truncated Shannon throughput, assuming 256 QAM as highest modulation format.

TABLE I. SIMULATION ASSUMPTIONS

| Channel Model | UMa 3D, according to 3GPP 36.873, [12] |
|---|---|
| Inter Site Distance | 500m |
| Nr. Cells | 7 tri-sector-cells, with wrap around |
| Scheduler | SUS [14], single layer transmission per scheduled UE |
| BS Antenna Configuration | various shapes, physical elements with 60 deg horizontal and vertical beam width |
| Number of Antenna Ports | 64/32/16, according to array shape |
| BS TX Power | 43 dBm (256 elements), 40 dBm (128 elements), 37 dBm (64 elements) |
| BS Ant. Downtilt | 12 deg mechanical downtilt |
| UE Antenna. | 1x1x2 (omnidirectional X-pol. antenna) |
| Max. UE TX PSD | 23 dBm/10 MHz |
| User Distribution | random placement, UEs are redropped until each sector serves an equal number of UEs |
| UL/DL Channel Measurements | UL: 1 sample per coherence block, UL/DL channel reciprocity assumed |
| Beamforming | Zero Forcing (ZF) |
| Channel Estimation | UL pilot contamination. <Nr. Users> orthogonal pilots per sector and coherence block, pilot reuse 3 between sectors, UL pilot power control |
| Link Adaptation | perfect link adaptation based on truncated Shannon |
| Traffic Model | full buffer |
| Bandwith | 10 MHz |
| Physical Layer | OFDM with 15 kHz subcarrier spacing and 14 OFDM symbols per subframe |
| Subframe Duration | 1ms including DL data and UL pilot overhead |
| Carrier Frequency | 2 GHz |
| UE Speed | 3 km/h |

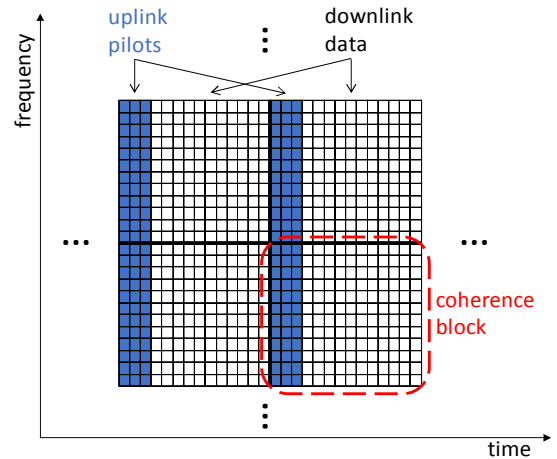

Fig. 4. TDD with reference symbols in UL and data transmission in DL, $W_c$ = coherence bandwidth, $T_c$ = coherence time

## B. Simulation results

The impact of the array shape on spectral efficiency is shown in Fig. 5 exemplarily using a zero-forcing precoder.

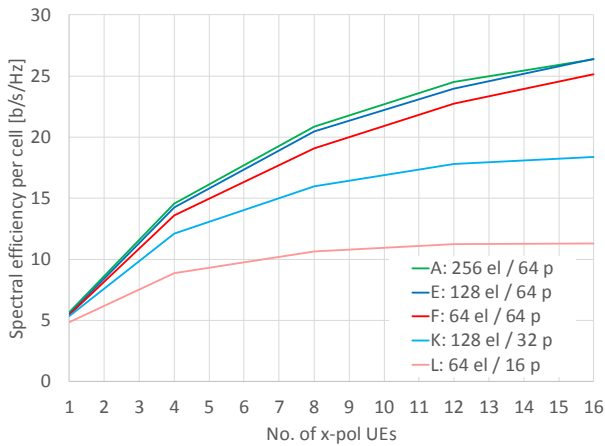

Fig. 5. Spectral efficiency of different no. of UEs for different array types (el = antenna element, p = antenna port)

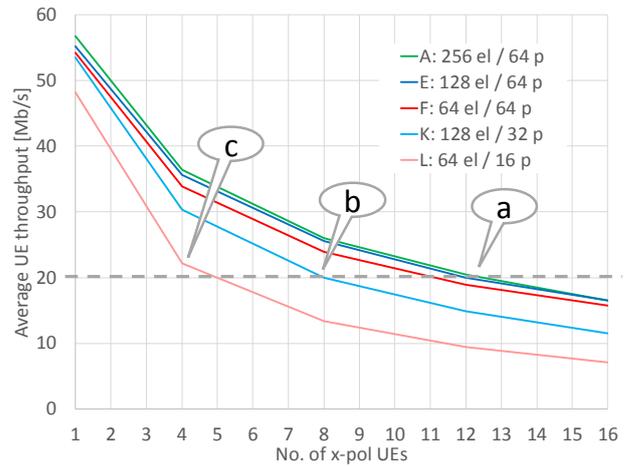

Fig. 6. Average UE throughput depending on no. of UEs and array type (el = antenna element, p = antenna port)

The spectral efficiency increases with the number of simultaneously served UEs for all array types. A, E and F have almost similar performance, because the system operates in interference limited condition and the array aperture remains the same. Also, the number of antenna ports remains. Only the antenna pattern per subarray changes in vertical direction, which has minor impact with the selected UMa-3D channel model because the angular spread in vertical direction is lower than in horizontal direction. Array type K and L achieve less increase in spectral efficiency with growing number of UEs. The reason is the lower number of ports and therefore less degrees of freedom for the MIMO precoder when the number of MIMO layers increases. In our scenario this leads to a saturation of spectral efficiency if $K \geq P/2$

It should be noted that for one up to four UEs the difference between K and A, E, F is rather low, and for only one active UE even array type L achieves comparable performance.

The analysis of the average UE throughput shown in Fig. 6 indicates that even with the reduction of the spectral efficiency due to smaller arrays the average throughput per UE can be maintained. If we consider e.g. the average throughput with 12 UEs (point a in Fig. 6) and then assume a decline to eight active UEs, the average throughput can be maintained with an array of size K (point b in Fig. 6). Further, when only four or less UEs are active, even with the small array type L (point c in Fig. 6) an even higher average throughput can be achieved.

These simulation results motivate the idea to consider an adaptation of the array size in case of low system load, to reduce the overall power consumption and increase energy efficiency, but maintaining the user throughput and satisfying the service needs. Therefore, for the considered array architecture a power consumption model has been derived and power consumption in relation to performance has been investigated for the different array types.

## IV. POWER CONSUMPTION

### A. Power consumption model

Globally, from power consumption perspective, a massive MIMO system can be split into the following two main fields: the baseband unit and the RF frontend unit. The latter one can again be split into the digital RF frontend unit - mainly controlling the RF frontends, if necessary comprising the analogue RF frontend close signal processing like digital predistortion and reduction of signal dynamic, interface processing as well as certain massive MIMO processing - and the analogue RF frontend, comprising transmit as well as receive signal conversion, amplification and filtering. Investigations and findings presented in this paper are based on the analogue RF frontend unit power consumption model. The power consumption of the baseband and the digital RF frontend unit is not considered here.

The analogue RF frontend power model used for our investigations comprises the main contributors from power consumption perspective (D/A- & A/D-converters, modulators, power amplifiers and low noise amplifiers, etc.) as well as from insertion loss perspective (filters, phase shifters, etc.) for both directions, downlink and uplink. Realistic values for the power consumption, energy efficiencies as well as insertion loss have been assigned to the different building blocks, based on experience. Based on this, the power consumption of the different analogue RF frontend building blocks (TX conversion unit, TX power amplifier, RX conversion unit and RX low noise amplifier) can be calculated and depicted individually, allowing to e.g. assess their particular contribution to the overall system power consumption budget for different massive MIMO port and antenna configurations. The overall analogue RF frontend power consumption can globally be calculated as follows, where $\tilde{P}_y^{TXconv}$ and $\tilde{P}_y^{RXconv}$ describe the respective conversion unit power consumption of port y for downlink and uplink and $\tilde{P}_z^{PA}$ as well as $\tilde{P}_z^{LNA}$ describe the power consumption of the respective power amplifier z (downlink) and the respective low noise amplifier z (uplink):

$$\tilde{P}_{tot} = \sum_{y=1}^{P}\left(\tilde{P}_{y}^{TXconv} + \tilde{P}_{y}^{RXconv}\right) + \sum_{z=1}^{N}(\tilde{P}_{z}^{PA} + \tilde{P}_{z}^{LNA}) \quad (1)$$

For our investigations, for all conversion units y=1 to P as well as for all amplifiers z=1 to N the same power consumption has been assumed.

In case of hybrid massive MIMO where the number of ports P is less than the number of antennas N, we assumed that the power splitting in downlink direction is done before the power amplifier, thus the number of power amplifiers is equal to the number of antennas. In uplink direction the power combining is done after the low noise amplifier, thus again the number of low noise amplifiers is equal to the number of antennas. In both directions realistic splitting losses have been considered per antenna.

For the downlink power amplifier we have selected the Doherty amplifier concept for our investigations, constituting an energy efficient state-of-the-art concept. In [15] a GaN-technology based Doherty amplifier MMIC for massive MIMO application has been published, providing high energy efficiency and high gain. From this paper we extracted the efficiency values of the asymmetric Doherty for 8dB (full load), 11 dB (50% load) and 14 dB (25% load) back off as basis for our investigations of the analogue RF frontend power consumption for different antenna configurations and the previously mentioned different load situations. Since for our analogue massive MIMO frontend a higher gain was required, we assumed an additional pre-amplifier adding estimated 0,5W additional power consumption (assumed same for each of the three load scenarios) to the overall amplifier line-up consumption. TABLE II. shows from the left column to right column the back off, the related power added efficiency values extracted from [15] and the assumed power amplifier line-up efficiency when adding the 0,5W power consumption considering an additional low power preamplifier.

TABLE II. EXEMPLARY DOHERTY BASED AMPLIFIER LINE-UP EFFICIENCIES

| Back off (LTE) | Doherty MMIC PAE extracted from [15] | Assumed used Doherty based PA line-up PAE |
|---|---|---|
| 8dB (full load) | 43% | 40% |
| 11dB (50% load) | 32% | 29% |
| 14dB (25% load) | 21% | 18,5% |

B. *Power consumption and power efficiency analysis*

The power consumption analysis in Fig. 7 covers the contributions of conversion unit and LNA on receiver side, and conversion unit and PA on transmitter side. Furthermore, main losses, e.g. for filter as well as splitter and phase shifter in case of hybrid massive MIMO are considered. We assumed a system with 100 MHz bandwidth and a total transmit power of 53 dBm. Reference system is array type A with 256 elements and 64 ports. With a back off at full load of 7.8 dB TABLE II. gives an exemplary power efficiency of about 40% (Doherty). For a TDD duty cycle of 75% this leads to a PA power consumption per antenna element of 2.07 W. The individual contributions per antenna element of the other functional blocks are depicted in TABLE III.

TABLE III. EXEMPLARY POWER CONSUMPTION PER FUNCTIONAL BLOCK

| Functional block (parameter) | Power consumption [W] | Scaling per |
|---|---|---|
| PA ($P^{PA}$) @back off = 7.8dB | 2.07 | Antenna element |
| LNA ($P^{LNA}$) | 0.28 | Antenna element |
| Tx conversion ($P^{TXconv}$) | 2.34 | Antenna port |
| Rx conversion ($P^{RXconv}$) | 0.9 | Antenna port |

Using (1) we have calculated the contribution of these functional blocks to the overall power consumption for different loads in terms of number of UEs. The reference case is the fully loaded scenario where 16 UEs are served with array type A. Now, if we assume a reduced load, represented by eight UEs, we can consider three cases: using array type A, array type E and array type K, with a total transmit power reduced by 3 dB. Array type E and K have both 128 elements, but different number of ports (64 and 32, respectively).

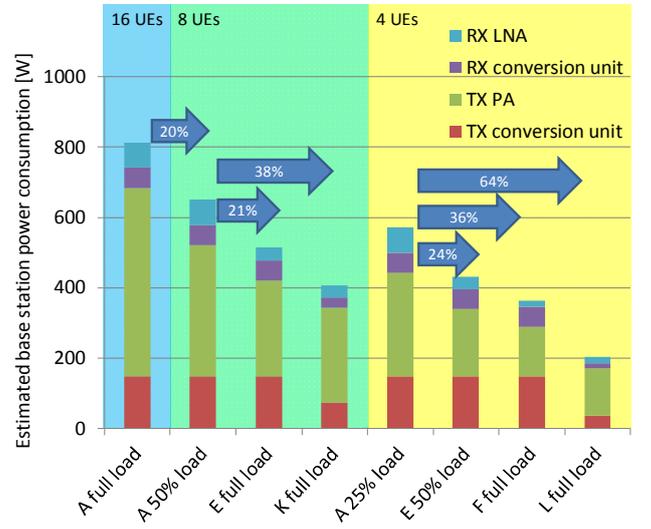

Fig. 7. Power consumption of different array architectures for different traffic load

Although the setup with array type A ("A 50% load" in Fig. 7) already consumes 20% less power than with full load, the array types E and K save additional 21% and 38% of power. When further reducing the load and serving only four UEs (at a total transmit power reduced by 6 dB), the array types F and L save further a significant amount of power compared to array type A with 25% load ("A 25% load" in Fig. 7). Even array type E with 50% load ("E 50% load" in Fig. 7) shows lower power consumption than array type A. In all cases of the comparison, the remaining transmit power per UE is the same as with 16 UEs. The considered scenarios match with those used for the performance analysis. So, we can state that in case of a lower number of UEs in the system the array size and related overall power consumption can be reduced without compromising the average throughput per UE.

To complete the picture now we calculate the energy efficiency in terms of power consumption per UE per Mb/s. The evaluation of the spectral efficiency for different array types and number of UEs leads to the results in Fig. 8. Using the full load scenario with array type A as 100%, we see that the power efficiency improves if the array size is adapted to the load. An extreme scenario, which shows the need for array adaptation, is the case of four UEs served with array type A. Although the total transmit power is only 25% of the full load scenario, here the cost in terms of power per UE per Mb/s is even increased compared to the full load scenario.

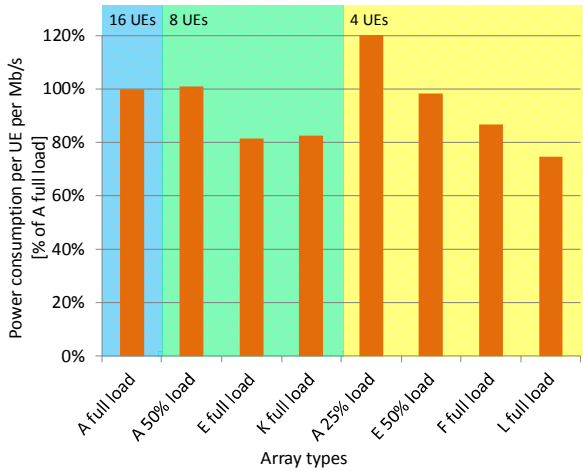

Fig. 8. Power consumption per UE per Mbit/s for different array architectures and traffic load

## V. CONCLUSION AND OUTLOOK

This paper presents a solution for energy efficient operation of massive MIMO systems in practical scenarios with strongly varying user densities. Adaptation of the antenna array configuration according to the traffic load can reduce power consumption. Different array configurations have been analyzed using a power consumption model of the analog RF frontend, which is the most dominant contribution to overall power consumption.

The results show that an adaptation of the array, i.e. switching off antenna elements in case of low number of users can significantly improve the energy efficiency of such a system, while maintaining the per UE performance. So, in case of e.g. only 4 simultaneous users an adaptation from the full-size array type A to the smaller array type F saves about 55% of power but reduces the spectral efficiency only about 10%. Since the number of MIMO layers is also reduced, this means that the average throughput per user can be maintained.

For the adaptation to a low number of active UEs it is essential to reduce the array size in the appropriate way to maintain performance. It depends on the channel characteristic, whether a size reduction in vertical or horizontal direction is more beneficial. Also, the array architecture and the number of ports versus the number of MIMO layers has impact. Switching of elements within a subarray performs better than the reduction of complete subarrays. However, the energy efficiency for array type L compared to F is slightly better.

Future studies will therefore investigate optimized dynamic adaptation schemes also with respect to feasibility within 5G NR standard and with respect to specific service type requirements. In addition, also power consumption of further functional blocks e.g. related to baseband processing, digital predistortion and further architecture variants will be considered.

ACKNOWLEDGMENT

This work has been performed in the framework of the Horizon 2020 project ONE5G (ICT-760809) receiving funds from the European Union. The authors would like to acknowledge the contributions of their colleagues in the project, although the views expressed in this contribution are those of the authors and do not necessarily represent the project.

REFERENCES

[1] T. L. Marzetta, "Noncooperative Cellular Wireless with Unlimited Numbers of Base Station Antennas," in *IEEE Transactions on Wireless Communications*, vol. 9, no. 11, pp. 3590-3600, November 2010.

[2] E. G. Larsson, O. Edfors, F. Tufvesson and T. L. Marzetta, "Massive MIMO for next generation wireless systems," in *IEEE Communications Magazine*, vol. 52, no. 2, pp. 186-195, February 2014.

[3] J. Hoydis, S. ten Brink and M. Debbah, "Massive MIMO in the UL/DL of Cellular Networks: How Many Antennas Do We Need?," in *IEEE Journal on Selected Areas in Communications*, vol. 31, no. 2, pp. 160-171, February 2013.

[4] E. Björnson, J. Hoydis, M. Kountouris and M. Debbah, "Massive MIMO Systems With Non-Ideal Hardware: Energy Efficiency, Estimation, and Capacity Limits," in *IEEE Transactions on Information Theory*, vol. 60, no. 11, pp. 7112-7139, Nov. 2014.

[5] E. Björnson, L. Sanguinetti, J. Hoydis and M. Debbah, "Optimal Design of Energy-Efficient Multi-User MIMO Systems: Is Massive MIMO the Answer?," in *IEEE Transactions on Wireless Communications*, vol. 14, no. 6, pp. 3059-3075, June 2015.

[6] M. Feng, S. Mao and T. Jiang, "BOOST: Base station ON-OFF switching strategy for energy efficient massive MIMO HetNets," *IEEE INFOCOM 2016 - The 35th Annual IEEE International Conference on Computer Communications*, San Francisco, CA, 2016, pp. 1-9.

[7] M. Olyaee, M. Eslami and J. Haghighat, "An energy-efficient joint antenna and user selection algorithm for multi-user massive MIMO downlink," in *IET Communications*, vol. 12, no. 3, pp. 255-260, 2 20 2018.

[8] Z. Liu, W. Du and D. Sun, "Energy and Spectral Efficiency Tradeoff for Massive MIMO Systems With Transmit Antenna Selection," in *IEEE Transactions on Vehicular Technology*, vol. 66, no. 5, pp. 4453-4457, May 2017.

[9] H. Li *et al*., "Energy Efficient Antenna Selection Scheme for Downlink Massive MIMO Systems," *2018 IEEE International Symposium on Circuits and Systems (ISCAS)*, Florence, Italy, 2018, pp. 1-4.

[10] Wenjia Liu, Shengqian Han, Chenyang Yang, "Is Massive MIMO Energy Efficient?", arXiv:1505.07187v1 [cs.IT], 2015

[11] A. F. Molisch *et al*., "Hybrid Beamforming for Massive MIMO: A Survey," in *IEEE Communications Magazine*, vol. 55, no. 9, pp. 134-141, 2017.

[12] 3GPP TR 36.873, V12.7.0, "Study on 3D channel model for LTE (Release 12)", 3rd Generation Partnership Project (3GPP), Dec. 2017.

[13] 3GPP TS 36.211, V15.1.0, "Physical channels and modulation (Release 15)", 3rd Generation Partnership Project (3GPP), Mar. 2018

[14] Taesang Yoo and A. Goldsmith, "On the optimality of multiantenna broadcast scheduling using zero-forcing beamforming," in *IEEE Journal on Selected Areas in Communications*, vol. 24, no. 3, pp. 528-541, March 2006.

[15] Stefan Maroldt, Mariano Ercoli, "3.5-GHz Ultra-Compact GaN Class-E Integrated Doherty MMIC PA for 5G massive-MIMO Base Station Applications", Proceedings of the 12[th] European Microwave Integrated Circuits Conference, pp. 196-199, Nuremberg 9[th] - 10[th] October 2017